\documentclass[5p]{elsarticle} % replaces elsart for hyperref etc.
\usepackage{hyperref}
\hypersetup{colorlinks=true, linkcolor=blue, filecolor=blue, pagecolor=blue, urlcolor=blue}
\usepackage{graphics}
\usepackage[normalem]{ulem} %emphasize further-on italic 
\usepackage{colordvi}% for using colordvi:

\usepackage{times}
\usepackage{epsfig}
\usepackage{cite}
\usepackage{amsmath}
\usepackage{pstricks}
\usepackage{pst-node}

\def \eps {\varepsilon}

\newcommand{\be}{\begin{equation}}
\newcommand{\ee}{\end{equation}}
\newcommand{\bea}{\begin{eqnarray}}
\newcommand{\eea}{\end{eqnarray}}
\newcommand{\non}{\nonumber}
\newcommand{\nl}{\nonumber \\}

\newcommand{\nn}{\nonumber}

\begin{document}
%\allowdisplaybreaks
%\sloppy

\begin{frontmatter}
  \title{%
 {\small \texttt{DESY 09-101}\hfill $\phantom{.}$\\}
 {\small \texttt{BI-TP 2009/15}\hfill $\phantom{.}$} 
 \\
 { \small \texttt{HEPTOOLS 09-020}\hfill $\phantom{.}$}
 \\
 {\small \texttt{SFB/CPP-09-63} \hfill$\phantom{.}$} 
%\hfill 
\\
A recursive reduction of tensor Feynman integrals 
}
\author[desy]{Th.~Dia\-ko\-ni\-dis}
\ead{Theodoros.Diakonidis@desy.de}
\author[bielefeld]{J. Fleischer}
\ead{Fleischer@physik.uni-bielefeld.de}
\author[desy]{T. Riemann\corref{cor1}}
\ead{Tord.Riemann@desy.de}
\author[desy]{J. B. Tausk}
\ead{Bas.Tausk@desy.de}
\address[desy]{Deutsches Elektronen-Synchrotron, DESY, 
   Platanenallee 6, 15738 Zeuthen, Germany}
\address[bielefeld]{Fakult\"at f\"ur Physik, Universit\"at Bielefeld, Universit\"atsstr. 25,  33615
Bielefeld, Germany
}
\cortext[cor1]{Corresponding author}
\begin{abstract}
We perform a new, recursive reduction of one-loop $n$-point rank $R$ tensor Feynman integrals [in short: $(n,R)$-integrals] for $n\leq 6$ with $R\leq n$ by representing  $(n,R)$-integrals in terms of    $(n,R-1)$- and $(n-1,R-1)$-integrals.
We use the known representation of tensor integrals in terms of
scalar integrals in higher dimension, which are then reduced by
recurrence relations to integrals in generic dimension. 
With a systematic
application of metric tensor representations in terms of chords,
and by decomposing and recombining these representations, a recursive reduction for the 
tensors is found.
The procedure represents a compact, sequential algorithm for numerical evaluations of tensor Feynman integrals  appearing in next-to-leading order contributions to massless and massive three- and four- particle production at LHC and ILC, as well as at meson factories.
\end{abstract}


\end{frontmatter}

%----------------------------------------------------------------------------------------
\section{Introduction}
For the evaluation of next-to-leading order contributions to processes at high energy colliders like LHC and ILC, but also at meson factories, one needs an efficient and reliable treatment of
$(n,R)$-integrals, i.e.
 $n$-point Feynman integrals with tensor rank $R$.
Typically $n\leq 6$ and $R\leq n$ may be needed for final states with massive particles.
For  $n\leq 4$ the Passarino-Veltman reduction \citep{Passarino:1978jh} may be  applied.
 For $n=5,6$ there are a variety of reduction schemes; for an overview, see e.g. \citep{Weinzierl:2007vk,Bern:2008ef,Binoth:2009fk}. 
In a recent article \citep{Diakonidis:2008ij}, we derived such a tensor reduction scheme for pentagons and hexagons using the Davydychev-Tarasov approach \citep{Davydychev:1991va,Tarasov:1996br} for tensors of rank $R\leq 3$.
Recurrence relations to reduce dimensions and indices have been applied 
with  systematic use of
 \emph{signed minors} \citep{Melrose:1965kb} as described in \citep{Fleischer:1999hq}.
Simplifications
were derived in \citep{Diakonidis:2008ij} for $n =5, R=2,3$ and $n=6, R=2,3,4$.
A corresponding numerical   {Mathematica} package   {hexagon.m} is publicly available 
{at  http://prac.us.edu.pl/$\sim$gluza/hexagon/ }
\citep{Diakonidis:2008dt}.

{In this article,  we present a new, recursive tensor reduction 
which 
extends the reduction
up to  $(6,6)$- and $(5,5)$-integrals.
In this reduction tensor integrals with $n\leq 4$ occur which can also be dealt with in a recursive manner.
This will be sufficient to evaluate the one-loop amplitudes of four-particle production at LHC and ILC.
The new reductions 
rest on a new master formula, equation  (\ref{tensor5general}) for five-point functions and corresponding ones for simpler functions.
In principle, the tensor integrals  with $n \leq 4$ might be treated following \citep{Passarino:1978jh}, e.g. with the Fortran package LoopTools/FF 
\citep{Hahn:1998yk2,vanOldenborgh:1990yc}.
LoopTools treats loops with \emph{massive} propagators for $n \leq 5$,%
\footnote{We observed problems in certain configurations with light-like external particles.}
 and Golem \citep{Binoth:2008uq} with \emph{massless} propagators for $n \leq 6$.
Unfortunately, there is no publicly available numerical Fortran package with a stable treatment of \emph{both} massive and massless particles in the loop.
In this situation, it appears natural to work out the complete reduction scheme
for the whole chain of tensors in a systematic way.

The tensor reductions given for  $n \leq 4$ in \citep{Passarino:1978jh} 
express the tensors in terms of scalar 1- to 4-point functions.
For tensor 5-point functions, reductions to tensor 4-point functions with rank less by one
have been presented in \citep{Binoth:2005ff} and \citep{Denner:2002ii,Denner:2005nn}, and also in \citep{delAguila:2004nf,vanHameren:2009vq} tensor recursions are discussed.
A representation of scalar $(N+1)$-point functions (including integrals with powers of the loop momentum in the numerator) in terms of $N$-point functions in $N$ integer dimensions was derived in \citep{vanNeerven:1983vr}.
For $N=4$, such a representation was derived already in \citep{Melrose:1965kb}.
The general case of tensor integrals using dimensional regularization was treated in a series of papers \citep{Bern:1992em,Bern:1993kr,Binoth:1999sp,Duplancic:2003tv}, thereby allowing also for massless particles.

Our reductions express
   $(n,R)$-integrals recursively in terms of    $(n,R-1)$- and $(n-1,R-1)$-integrals
for $n = 2 \dots 6$.
Although all approaches have identical basis elements and thus have to have equivalent tensor coefficients when compared after complete reduction, we would like to stress that they 
allow for quite  different algorithmic realizations.
}

  {The article is organized as follows.
Basic formulae are introduced in Section \ref{sec-basics}.
Section \ref{sec-5r} contains the main result, the recursive tensor reduction, based on  master representations for the $(5,R)$-integrals.
As a demonstration, we derive the $(5,4)$-integrals in more detail in Section \ref{sec-rank4}.
Section  \ref{sec-6r} is a short comment on the $(6,R)$-integral reductions.
In Section \ref{sec-discussion} we state some properties of
the auxiliary vectors used in the recursions,
which also allow for an alternative and simple derivation of our master formula.
We finish with a short Summary.}
%--------------------------------------------------------------------
\section{Basic formulae\label{sec-basics}}
%--------------------------------------------------------------------
We study Feynman tensor integrals in the \emph{generic dimension} $d=4-2\epsilon$ with $n$ external legs:
\begin{eqnarray}\label{definition}
 I_n^{\mu_1\cdots\mu_R} &=&  ~~\int \frac{d^d k}{i\pi^{d/2}}~~\frac{\prod_{r=1}^{R} k^{\mu_r}}{\prod_{j=1}^{n}c_j^{\nu_j}},
\end{eqnarray}
where the denominators $c_j$ have \emph{indices} $\nu_j$ and \emph{chords}
$q_j$:
\begin{eqnarray}\label{propagators}
c_j &=& (k-q_j)^2-m_j^2 +i\varepsilon .
\end{eqnarray}
We will assume $\nu_j=1$ in the following, but a generalization of the results to arbitrary indices is straightforward. 

The iteration of reduction steps will be performed until the level of $(n,0)$-integrals with $n\leq 4$ is reached.
% or, numerically, going upward to increasing ranks of the tensors. 
In this chain, the following well-known scalar reductions \citep{Fleischer:1999hq} are needed:
\begin{eqnarray}\label{scaln}
 I_{n}
&=&     \sum_{s=1}^{n}      
\frac{{0 \choose s}_n}
{{0\choose 0}_n}
I_{n-1}^{s}, ~~~~n=5,6.
\end{eqnarray}
The simplest, but typical tensor is the vector integral:
\bea\label{scalarred}
  I_n^{\mu} & =&  
-\sum_{i=1}^{n} \, q_i^{\mu} \, I_{n,i}^{[d+]} .
\eea
The integrals $I_{n,i}^{[d+]}$ are scalar $n$-point integrals, obtained from $I_{n}$ by raising the index  of line $i$ by one unit ($\nu_i=2$ then) and replacing the generic dimension by dimension $d+2$.%
\footnote{Analogously, $I_{n,ij}^{[d+]^2}$ has indices $\nu_i=2$ and $\nu_j=2$ and is defined in $d+4$ dimensions, etc.}
In a next step, we apply the  recursion relations derived in \citep{Tarasov:1996br,Fleischer:1999hq}
in order to eliminate the shifts of dimension and indices. 
The details of the derivations, which are relatively easy for lower rank tensors, 
get complicated, due to many cancellations, for tensors of higher ranks.

The recursion relation for the vector coefficients in (\ref{scalarred}) reads for $n\leq 5$:
\begin{eqnarray}
\label{A511}
I_{n,i}^{[d+]}&=&-\frac{{0\choose i}_n}{\left(  \right)_n} I_{n} +
 \sum_{s=1}^{n} \frac{{s\choose i}_n}{\left(  \right)_n} I_{n-1}^{s} . 
\end{eqnarray}
For $n=1$ the second term in (\ref{A511}) vanishes and for
$n>5$ the denominator in (\ref{A511}) is $\left(  \right)_n=0$.
The case $n=6$ is of practical importance and will be discussed below.
$I_{n}$ is the scalar integral, and $I_{n-1}^s$ the integral $I_{n}$ where line $s$ has been scratched.%
\footnote{Analogously, in $I_{n-2}^{\{\mu\},st}$  lines $s$ and $t$ have been scratched, etc.}
The objects like ${s\choose i}_n$ are \emph{signed minors}, and $\left(  \right)_n$ is the \emph{modified Cayley determinant}.
For explicit definitions, see \citep{Melrose:1965kb} or Appendix A of \citep{Diakonidis:2008ij}.%
\footnote{The \emph{Gram determinant} is 
$|2 q_jq_k|, ~ j,k=1,\cdots, n-1$, and for $q_n=0$ we have $\left(  \right)_n=-|2 q_jq_k|$.}
Thus, we can write for the vector $n$-point function [i.e. the $(n,1)$-integral] (\ref{scalarred}) for $n\leq 5$ :
\bea\label{vector}
I_n^{\mu} &=&
 I_n Q_0^{\mu}-\sum_{s=1}^{n} I_{n-1}^s Q_s^{\mu},
\eea
In (\ref{vector}) we introduced the  \emph{auxiliary vectors} $Q_s^{\mu}$:
{
\bea
\label{Qs}
 Q_s^{\mu}&=&\sum_{i=1}^{n}  q_i^{\mu} \frac{{s\choose i}_n}{\left(  \right)_n},~~~ s=0, \dots, n.
\eea
}
Vectors (\ref{Qs}) are \emph{universal} and will appear in more involved reductions again.
Indeed, (\ref{vector}) is what we want to {obtain}  further on, i.e. we will look for analogous relations for  higher rank tensors in the following.

Equation (\ref{vector}) is essentially due to recursion relation (30) of \citep{Fleischer:1999hq}, which reduces simultaneously dimension and index (let us call it \emph{type I recursion}). 
For $(n,R)$-integrals with rank $R\geq 2$, a complication arises due to the appearence
of the $g^{\mu \nu}$-tensor in the reduction to scalar functions, as may be seen from the simplest case of an $(n,2)$-integral:
\bea
\label{vectorred}
 I_{n}^{\mu\, \nu}
 &=&
 \sum_{i,j=1}^{n} \, q_i^{\mu}\, q_j^{\nu} \, {\nu}_{ij} \,  \, I_{n,ij}^{[d+]^2}
 -\frac{1}{2}   \, g^{\mu \nu}  \, I_{n}^{[d+]}  ,
 \eea
with ${\nu}_{ij}=1+\delta_{ij}$.
In $d=4$ dimensions, one may eliminate  $g^{\mu \nu}$ by  expressing it in terms of 
the $n$ different chords of the integral:
\bea \label{gmunu6}
g^{\mu \nu}&=&2 \sum_{i,j=1}^{6} q_i^{\mu} q_j^{\nu}\frac{{0i\choose 0j}_6}{{0\choose 0}_6},
\\\label{gmunun5}
g^{\mu \nu}&=&2 \sum_{i,j=1}^{5} q_i^{\mu} q_j^{\nu}\frac{{i\choose j}_5}{\left(  \right)_5},
\\\label{gmunun4}
 g^{\mu \nu} &=& 2\sum_{i,j=1}^{4} q_i^{\mu}  q_j^{\nu} \frac{{i\choose j}_4}{\left(  \right)_4}
+
\frac{8v^{\mu} v^{\nu}}{\left(  \right)_4},
\\\label{gmunun3}
g^{\mu \nu}&=&2 \sum_{i,j=1}^{3} q_i^{\mu} q_j^{\nu}\frac{{i\choose j}_3}{\left(  \right)_3}
+ \frac{4 v^{\mu\lambda} v^{\nu}_{~\lambda}}{\left(  \right)_3} .
\eea
For $n=6,5$, see \citep{Fleischer:1999hq}.
For $n<5$, we have to introduce extra terms \citep{vanOldenborgh:1989wn}, defined with the aid of: 
\bea\label{eqa2}
v^{\mu}&=&\eps^{\mu\lambda\rho\sigma}
          (q_1-q_4)_{\lambda} (q_2-q_4)_{\rho} (q_3-q_4)_{\sigma},
\\
\label{eqa3}
v^{\mu\lambda}&=&\eps^{\mu\lambda\rho\sigma}
     (q_1-q_3)_{\rho} (q_2-q_3)_{\sigma},
\eea
where 
\bea \label{eq02}
v^{\mu\lambda} v^{\nu}_{~\lambda}
&=&q_1^2 q_2^{\mu} q_2^{\nu}+q_2^2 q_1^{\mu} q_1^{\nu}-
(q_1 q_2) (q_1^{\mu} q_2^{\nu}+q_1^{\nu} q_2^{\mu})
\nl
&&~-
\left[ q_1^2q_2^2-(q_1  q_2)^2 \right] g^{\mu \nu} .
\eea
In (\ref{eq02}), the $q_1, q_2$ are short for the two 4-vectors in (\ref{eqa3}), $(q_i-q_3); i=1,2$.
We just mention that in our conventions $\textrm{det}\left(g^{\mu \nu}\right) = -1$ and 
$\eps^{0123}=-\eps_{0123}=+1$. 
Further, $v^2 = \frac{1}{8} ()_4 $.
{It is also interesting to note that the contractions of the sums appearing in (\ref{gmunun4}) and (\ref{gmunun3}) with their corresponding extra terms vanish, i.e. these terms are orthogonal.}

Now, we are ready to derive a systematic recursion algorithm.
%\newpage
%--------------------------------------------------------------------
\section{The $(5,R)$-integrals\label{sec-5r}}
%--------------------------------------------------------------------
The scalar $(5,0)$-integral is given in (\ref{scaln}), and the vector $(5,1)$-integrals in {(\ref{vector})}.
Applying recursion relations, one may derive the following master formula for the $(5,R)$-integrals:
\bea
I_5^{\mu_1  \dots \mu_{R-1} \mu}  =I_5^{\mu_1  \dots \mu_{R-1}} Q_0^{\mu} -  \sum_{s=1}^{5}
I_4^{\mu_1  \dots \mu_{R-1},s } Q_s^{\mu}. 
\label{tensor5general}
\eea
The formula is a generalization of (\ref{vector}). 
Equation (16) is given implicitly in \citep{Diakonidis:2008ij}, for $n=5, R\leq 3$:
by equation (3.7) for $R=2$ and by equation (3.19) for $R=3$, both in combination with equations (2.1) and (2.2) (these both for  $n=4$).
For $R=4$ see also section 4 and, alternatively, see also section 6. 

As a consequence, (\ref{tensor5general}) has a completely new tensor structure compared to what one is used to.
The explicit evaluation of (\ref{tensor5general}) will be discussed below for $R=3,4,5$.
The $(5,2)$-integrals follow immediately from (\ref{tensor5general}) since the vector integrals $I_5^\mu$ and $I_4^{\mu,s}$ on the r.h.s. are known from  (\ref{vector}).
%---------------------------------------------------------------------------
\subsection{The $(5,3)$-integral recursion family\label{sec-53}}
The master formula (\ref{tensor5general}) for $(5,3)$-integrals has on the r.h.s $(5,2)$- and $(4,2)$-integrals:
\bea
I_5^{\mu_1 \mu_{2} \mu}  =I_5^{\mu_1 \mu_{2}} Q_0^{\mu} -  \sum_{s=1}^{5}
I_4^{\mu_1 \mu_{2},s } Q_s^{\mu}. 
\label{tensor52a}
\eea
The $(5,2)$-integrals have already been discussed and may be expressed
by $(5,1)$- and $(4,1)$-integrals according to (\ref{tensor5general})
and (\ref{vector}).
For the $(4,2)$-integrals, we need now a reduction analogous to the master formula (\ref{tensor5general}) and may use as a starting point 
(\ref{vectorred}).
The dimensional shifts in (\ref{vectorred}) may be treated with a reduction of type I:
\begin{eqnarray}
\label{A422}
{\nu}_{ij} I_{4,ij}^{[d+]^2,s}
&=& -\frac{{0s\choose js}_5}{{s\choose s}_5} 
I_{4,i}^{[d+],s} +
 \sum_{t=1,\ne i}^{5} \frac{{ts\choose js}_5}{{s\choose s}_5} 
I_{3,i}^{[d+],st}
\nl
&&~+\frac{{is\choose js}_5}{{s\choose s}_5} I_{4}^{[d+],s} ,
%~~~(46!). 
\end{eqnarray}
and 
another one, needed for the reduction of the dimension only (not of an
index), let us call it reduction of \emph{type II}.
Equation (31) of \citep{Fleischer:1999hq} yields:
\begin{eqnarray}
\label{A401}
I_{4}^{[d+],s}&=& \frac{{0s\choose 0s}_5}{{s\choose s}_5} I_{4}^{s}
-
 \sum_{t=1}^{5} \frac{{ts\choose 0s}_5}{{s\choose s}_5} I_{3}^{st}
 .
\end{eqnarray}
After eliminating the $g^{\mu\nu}$ in (\ref{vectorred}) with the aid of (\ref{gmunun4}), we obtain:
\begin{equation}
\label{tensorshort}
I_4^{\mu\nu,s}  = I_4^{\mu,s} Q_0^{s,\nu} 
- \sum_{t=1}^{4} I_3^{\mu ,st } Q_t^{s,\nu}
- \frac{4v^{s,\mu} v^{s,\nu}}{\left( \right)_4} I_4^{[d+],s}.
\end{equation}
The notations $Q_{t}^{s,\nu}$ and $v^{s,\nu}$ mean that from the five chords of the $5$-point function
the chord $s$ is excluded, such that these vectors are constructed from four chords - as
given in (\ref{gmunun4}). 
In fact, (\ref{tensorshort}) together with  (\ref{A401}) is the reduction of the $(4,2)$-integrals, when it is combined with (\ref{vector}) and:
\bea
\label{3vector}
I_3^{\mu ,st}  &=& I_3^{st} \sum_{i=1}^{5} q_i^{\mu} \frac{{0st\choose ist}_5}{{st\choose st}_5}-
 \sum_{u=1}^{5} I_{2}^{stu} \sum_{i=1}^{5} q_i^{\mu} \frac{{ust\choose ist}_5}{{st\choose st}_5} 
\nl
&\equiv&   I_3^{st}  Q_0^{st,\mu}-\sum_{u=1}^{5} I_2^{ust} Q_u^{st,\mu},
\eea
where the upper indices $s,t$ in the $Q^{st}$-vectors are again introduced for the scratched lines. 
Observe
that for $i=s,t$ and $u=s,t$ there are no contributions so that indeed the indices are running only over
three values and objects like ${0st\choose ist}_5$ can indeed be read as ${0\choose i}_3$. In this
way (\ref{3vector}) is consistent with (\ref{vector}). If one is interested in $4$-point
functions from the very beginning one avoids of course this clumsy notation - but for the
present purpose of reducing $5$-point tensors to scalars it appears adequate to demonstrate this at least once.

This completes the $(5,3)$-, $(4,2)$-, and  $(3,1)$-integral recursions.
%---------------------------------------------------------------------
\subsection{The $(5,4)$-integral recursion family\label{sec-54}}
For the higher tensors of the
$5$-point function we need correspondingly higher tensors of the $4$-point functions, and the corresponding extra terms related to the elimination of the $g^{\mu\nu}$ have to be derived.

Thus, as a next step we seek a representation for the $(4,3)$-integrals 
which is needed for the $(5,4)$-integral recursion: 
\bea
I_5^{\mu_1 \mu_{2} \mu_{3} \mu}  =I_5^{\mu_1 \mu_{2}\mu_{3} } Q_0^{\mu} -  \sum_{s=1}^{5}
I_4^{\mu_1 \mu_{2}\mu_{3},s } Q_s^{\mu},
\label{tensor52}
\eea
but is  also of interest in its own. 
In the following, for the ease of notation, we will drop the scratches of line $s$.

A systematic application of recursions of \emph{type I} results in:
\bea
I_4^{\mu \nu \lambda}  
&=&
I_4^{\mu \nu} Q_0^{\lambda} -  \sum_{t=1}^{4}
I_3^{\mu \nu ,t } Q_t^{\lambda}-G^{\mu \lambda }I_4^{\nu,[d+]}
\nl
&& -~ G^{\nu \lambda }I_4^{\mu,[d+]} ,
\label{tensor43}
\eea
with (see also (\ref{gmunun4})):
\bea
\label{Gml}
G^{\mu \lambda}=\frac{1}{2} g^{\mu \lambda}-
\sum_{i,j=1}^{4} q_i^{\mu}  q_i^{\lambda} \frac{{i\choose j}_4}{\left(  \right)_4}
=  \frac{4v^{\mu} v^{\nu}}{\left(  \right)_4},
\eea
and
\bea
I_4^{\mu,[d+]}
&=& - \sum_{k=1}^{4} q_k^{\mu} I_{4,k}^{[d+]^2}
\nl
&=&
I_4^{[d+]} Q_0^{\mu} - \sum_{t=1}^{4} I_3^{[d+],t} Q_t^{\mu}.
\label{GV} 
\eea
In $I_4^{\mu,[d+]}$, besides $I_4^{[d+]}$ (known from (\ref{A401})), also
$I_3^{[d+],t}$ enters.
It may be reduced by a recursion of  \emph{type II}
quite similar to (\ref{A401}):
\begin{eqnarray}
 I_{3  }^{[d+],t}&=& \left[
\frac{{0t\choose 0t}_4}{{t\choose t}_4}I_{3  }^{t}-
\sum_{u=1}^4\frac{ {ut\choose 0t}_4}{{t\choose t}_4} I_{2  }^{tu} \right]{\frac{1}{d-2}} .
\label{A301}
\end{eqnarray}
Finally, our representation (\ref{tensor43}) of $I_4^{\mu \nu \lambda}$  contains the integrals $I_3^{\mu \nu ,t }$ and we have to reduce {them} also. 
Application of recursion relations yields {the analogue of (\ref{tensorshort}) for $n=4$}:
{
\begin{equation}\label{tensor32}
I_3^{\mu \nu ,t }
=
I_3^{\mu,t } Q_{0}^{t,\nu} -  \sum_{u=1}^{4} I_{2}^{\mu , tu}Q_{u}^{t,\nu}
-  
I_3^{[d+],t} 
\frac{2v^{t,\mu\lambda}v^{t,\nu}_{\lambda}}{{t\choose t}_4}.
\end{equation}
}
We made use of the definition (\ref{Qs}), which becomes here:
{
\bea
\label{Qst}
 Q_u^{t,\nu}&=&\sum_{i=1}^{4}  q_i^{\nu} \frac{{ut\choose it}_4}{ {t \choose t}_4},~~~ u=0, \dots, 4.
\eea
and
of the representation (\ref{gmunun3}) in order to express:}
\bea
G^{t,\mu \nu} &=& \frac{1}{2} g^{\mu \nu} -
\sum_{i,j=1}^{4} q_i^{\mu}  q_j^{\nu} \frac{{it\choose jt}_4}{{t\choose t}_4}
\nl
&=& \frac{2v^{t,\mu \lambda}v^{t,\nu}_{\lambda}}{{t\choose t}_4}. 
\label{Gt}
\eea

This completes the $(5,4)$-, $(4,3)$-, and  $(3,2)$-integral recursions.
%---------------------------------------------------------------------
\subsection{The $(5,5)$-integral recursion family\label{sec-55}}
%---------------------------------------------------------------------
For the tensor of rank $5$ of the $5$-point function we need further the tensor
of rank $4$ of the $4$-point function. 
\bea
I_5^{\mu_1 \mu_{2} \mu_{3} \mu_{4} \mu}  =I_5^{\mu_1 \mu_{2}\mu_{3}\mu_{4}  } Q_0^{\mu} -  \sum_{s=1}^{5}
I_4^{\mu_1 \mu_{2}\mu_{3}\mu_{4} ,s } Q_s^{\mu}.
\label{tensor52b}
\eea
Again, the systematic application of the recursion relations results in
\bea
\label{tensor44}
I_4^{\mu \nu \lambda \rho}  &=&I_4^{\mu \nu \lambda} Q_0^{\rho} -  \sum_{t=1}^{4}
I_3^{\mu \nu \lambda,t } Q_t^{\rho}
\\ \nonumber
&&~-
G^{\mu \rho } T^{\nu \lambda }- G^{\nu \rho } T^{\mu \lambda }-G^{\lambda \rho } T^{\mu \nu }
\eea
with
\bea \label{eq03}
T^{\mu \nu }=I_4^{\mu,[d+]} Q_0^{\nu} - \sum_{t=1}^{4} I_3^{\mu,[d+],t} ~ Q_t^{\nu}~-G^{\mu \nu }I_{4}^{[d+]^2} ,
\eea
where $G^{\mu \nu }$ and $I_4^{\mu,[d+]}$ are given in  (\ref{Gml}) and (\ref{GV}), respectively, and
\bea
I_3^{\mu,[d+],t}
&=&- \sum_{i=1}^{4} q_i^{\mu} I_{3,i}^{[d+]^2,t}
\nl
&=&I_3^{[d+],t} Q_0^{t,\mu} - \sum_{u=1}^{4} I_2^{[d+],tu} Q_u^{t,\mu}.
\label{Wt}
\eea
The $I_3^{[d+],t}$ is known from (\ref{A301}), 
{the $I_{4}^{[d+]}$ from (\ref{A401})}, 
and 
for completeness we specify also the other recursions of \emph{type II}:
\begin{equation}
\label{eq04}
I_{4}^{[d+]^2}=\left[\frac{{0\choose 0}_4}{\left(  \right)_4} I_{4}^{[d+]} -
 \sum_{t=1}^{4} \frac{{t\choose 0}_4}{\left(  \right)_4} I_{3}^{[d+],t} \right]{\frac{1}{d-1}},
\end{equation}
\begin{equation}
\label{eq05}
I_{2}^{[d+],tu}=\left[\frac{{0tu\choose 0tu}_4}{{tu\choose tu}_4} I_{2}^{tu}-
 \sum_{v=1}^{4} \frac{{0tu\choose vtu}_4}{{tu\choose tu}_4} I_{1}^{tuv} \right]{\frac{1}{d-1}}
.
\end{equation}
In (\ref{tensor44}) enters  also the $I_3^{\mu \nu \lambda,t }$ , which we evaluate to be:
\bea
I_3^{\mu \nu \lambda, t}  &=&I_3^{\mu \nu ,t} Q_0^{t,\lambda} -  \sum_{u=1}^{4}
I_2^{\mu \nu ,tu } Q_u^{t,\lambda}
\nl
&&~
- G^{t,\mu \lambda } I_3^{\nu,[d+],t} - G^{t,\nu \lambda } I_3^{\mu,[d+],t},
\label{tensor33}
\eea
where $G^{t,\mu \nu }$ is given in (\ref{Gt}) and   $I_3^{\mu,[d+],t}$ in (\ref{Wt}).

It remains to evaluate the tensor $I_2^{\mu \nu ,tu }$, for which we get:
\bea \label{eq06}
I_2^{\mu \nu ,tu }&=& 
I_2^{\mu ,tu } Q_0^{tu,\nu}
-\sum_{v=1}^{4}I_1^{\mu ,tuv }  Q_v^{tu,\nu}
\nl&&
-~
G^{tu,\mu \nu } I_2^{[d+],tu}.
\eea
The auxiliary vector $Q_v^{tu,\nu}$, defined analogously to (\ref{Qst}),
vanishes when $v=t$ or $v=u$. The sum over $v$ in (\ref{eq06}) therefore
consists of two terms, $v=i,i^{'}$, where $i < i^{'}$ and  $i,i^{'} \ne t,u$.
With {${tu\choose tu}_4=-2 (q_i-q_{i^{'}})^2 \equiv -2 q^2$} 
we have for the vector basis in (\ref{eq06}):
\bea
Q_0^{tu,\nu}&=&\frac{1}{2}(q_i+q_{i^{'}})^{\nu}
\nl
&&-~\frac{1}{2 q^2}(m_i^2-m_{i^{'}}^2)(q_i-q_{i^{'}})^{\nu},
\\
Q_i^{tu,\nu}&=& - Q_{i^{'}}^{tu,\nu}
          \; = \; \frac{1}{2 q^2} (q_i-q_{i^{'}})^{\nu} ,
\\
G^{tu,\mu\nu}&=&\frac{1}{2}\left(g^{\mu \nu}-\frac{(q_i-q_{i^{'}})^{\mu}(q_i-q_{i^{'}})^{\nu}}{q^2}
\right).
\eea
Finally, in (\ref{eq06}) appear the integrals:
\bea \label{eq07}
I_2^{\mu ,tu }&=&I_2^{tu } Q_0^{tu,\mu}-\sum_{v=1}^{4} I_1^{tuv }Q_v^{tu,\mu}
\\
&=&\frac{1}{2}I_2^{tu }(q_i+q_{i^{'}})^{\mu}
\non  
\\
&&-\frac{1}{2}(m_i^2-m_{i^{'}}^2)\frac{I_2^{tu }(q^2)-I_2^{tu }(0)}{q^2}(q_i-q_{i^{'}})^{\mu}\non 
\eea
and 
\bea\label{eq08}
I_1^{\mu,tuv }&=&-q_i^{\mu} I_{1,i}^{[d+],tuv}=
q_i I_1^{tuv}, i \ne t,u,v. \non  \\
\eea
The last equality in (\ref{eq08}) follows from:
\bea
I_{1,i}^{[d+],tuv}&=&-\frac{{0tuv\choose ituv}_4}{{tuv\choose tuv}_4} I_1^{tuv}=
-
I_1^{tuv}.
\eea

This completes the $(5,5)$-, $(4,4)$-,   $(3,3)$-, $(2,2)$- and $(1,1)$-integral recursions.
%-------------------------------------------------------------------ssh -l jfleis pub.ifh.de-
\section{Derivation of the master formula for   $(5,4)$-integrals\label{sec-rank4}}
%--------------------------------------------------------------------
%
As mentioned before, (\ref{tensor5general}) is implicitly contained in \citep{Diakonidis:2008ij}, but only up to $R=3$.
We will make use of it, however, up to $R=5$ at least.
To demonstrate the general approach of how to obtain this relation, we give details of its analytic proof for $R=4$.
In particular we observe huge cancellations of higher dimensional integrals in (2.4) of \citep{Diakonidis:2008ij}.
To start with, as usual we write this relation as:
\bea
I_{5}^{\mu\, \nu\, \lambda\, \rho}= \sum_{i,j,k,l=1}^{5} \, 
q_i^{\mu}\, q_j^{\nu} \, q_k^{\lambda} \, q_l^{\rho} I_{5,ijkl},
\label{tensor4}
\eea
where (\ref{gmunun5}) has to be used. 
The highest dimensional integral
occuring now is  $n_{ijkl} I_{5,ijkl}^{[d+]^4}$, with $n_{ijkl}={\nu}_{ij}{\nu}_{ijk}{\nu}_{ijkl}, {\nu}_{ijkl}= 1+\delta_{il}+\delta_{jl}+\delta_{kl}$ etc., for which we need the recursion relation: 
\bea
\label{A544}
{\nu}_{ijkl} I_{5,ijkl}^{[d+]^4} &=&
-\frac{{0\choose l}_5}{\left(  \right)_5} I_{5,ijk}^{[d+]^3} 
+~
\sum_{\substack{s=1 \\ s \neq i,j,k}}^5
\frac{{s\choose l}_5}{\left(  \right)_5} I_{4,ijk}^{[d+]^3,s} 
\nn\\
&&+~ 
 \frac{{i\choose l}_5}{\left(  \right)_5} I_{5,jk}^{[d+]^3}+
 \frac{{j\choose l}_5}{\left(  \right)_5} I_{5,ik}^{[d+]^3}
\nn\\ 
&&+~ \frac{{k\choose l}_5}{\left(  \right)_5} I_{5,ij}^{[d+]^3}.
\eea
In the sum of the last three terms of (\ref{A544}), to be abbreviated as $[ijk]^{(l)}$,
it is understood that there occur no equal indices among $i,j,k$ (otherwise it would be
contained as a 4-point function in the second term on the right-hand side of (\ref{A544})). 
With the remaining factor
${\nu}_{ij}{\nu}_{ijk}$ of the integral $n_{ijkl} I_{5,ijkl}^{[d+]^4}$ we can rewrite:
\begin{eqnarray}
\label{equal}
{\nu}_{ij}{\nu}_{ijk} [ijk]^{(l)}
&=&{\nu}_{jk}\frac{{i\choose l}_5}{\left(  \right)_5} I_{5,jk}^{[d+]^3}+
{\nu}_{ik}\frac{{j\choose l}_5}{\left(  \right)_5} I_{5,ik}^{[d+]^3}
\nn\\
&&+~{\nu}_{ij}\frac{{k\choose l}_5}{\left(  \right)_5} I_{5,ij}^{[d+]^3} ,
\label{observation}
\end{eqnarray}
and as a result we have:
\begin{gather}%\bea
I_{5,ijkl} =
{\nu}_{ij}{\nu}_{ijk} 
\left[-\frac{{0\choose l}_5}{\left(  \right)_5} I_{5,ijk}^{[d+]^3}+
\sum_{s=1,s \ne i,j,k}^{5} \frac{{s\choose l}_5}{\left(  \right)_5} I_{4,ijk}^{[d+]^3,s}\right]   \nn\\
-\left[
{\nu}_{kl}\frac{{i\choose j}_5}{\left(  \right)_5} I_{5,kl}^{[d+]^3}+
{\nu}_{jl}\frac{{i\choose k}_5}{\left(  \right)_5} I_{5,jl}^{[d+]^3}+
{\nu}_{il}\frac{{j\choose k}_5}{\left(  \right)_5} I_{5,il}^{[d+]^3} \right] \nn\\
+~~~\frac{1}{\left(  \right)_5^2}
\left[{i\choose j}_5 {k\choose l}_5 + {i\choose k}_5 {j\choose l}_5 + 
      {j\choose k}_5 {i\choose l}_5 \right] I_5^{[d+]^2},
\label{Iijkl}
\end{gather}%\eea
where (\ref{equal}) has already completely cancelled  against the last three
terms of the second sum of (2.4) in \citep{Diakonidis:2008ij}. 
With the  further recursion:
\bea
\label{A523}
{\nu}_{il} I_{5,il}^{[d+]^3}&=&-\frac{{0\choose l}_5}{\left(  \right)_5} I_{5,i}^{[d+]^2} +
 \sum_{\substack{s=1\\s \ne i}}^{5} 
\frac{{s\choose l}_5}{\left(  \right)_5} I_{4,i}^{[d+]^2,s} \nn\\
&&+ \frac{{i\choose l}_5}{\left(  \right)_5} I_{5}^{[d+]^2} ,
\eea
one observes that the $I_{5}^{[d+]^2}$ term cancels against the last row of (\ref{Iijkl})
and the remaining $I_5$ integrals combine to $I_{5,ijk}$ according to (3.18) of \citep{Diakonidis:2008ij}
with the result:
\begin{gather}
I_{5,ijkl}
=
\frac{{0\choose l}_5}{\left(  \right)_5} I_{5,ijk}
\nn\\
- 
\left[\frac{{i\choose j}_5}{\left(  \right)_5}\sum_{\substack{s=1 \\ s \neq k}}^{5}
\frac{ {s\choose l}_5}{\left(  \right)_5}I_{4,k}^{[d+]^2,s}
+
(i \leftrightarrow k) + (j \leftrightarrow k) \right]
\nn \\
+ {\nu}_{ij}{\nu}_{ijk} 
\sum_{\substack{s=1\\ s\neq i,j,k}}^{5} \frac{{s\choose l}_5}{\left(  \right)_5} I_{4,ijk}^{[d+]^3,s} 
\label{raw4}
\end{gather}
This proves our statement: The first term corresponds to the (5,3)-integral as claimed in
(\ref{tensor5general}). 
The rest, according to (2.3) of \citep{Diakonidis:2008ij},
corresponds to the (4,3)-integral, again corresponding to (\ref{tensor5general}).
A similar proof for $n=5, R=5$, equation (\ref{tensor52b}), is a bit more lengthy, but see also the derivation in section 6.
We would like to mention that a pedagogical introduction to the techniques applied may be found in \citep{Fleischer-calc:2009}.
%}
%--------------------------------------------------------------------

%--------------------------------------------------------------------
\section{The $(6,R)$-integrals\label{sec-6r}}
%--------------------------------------------------------------------
Representing the $g^{\mu \nu}$-tensor by (\ref{gmunu6}), one 
has the analogue of (\ref{tensor5general})  for the  $(6,R)$-integrals: 
\bea\label{tensor6general}
I_6^{\mu_1 \dots \mu_{R-1} \rho}  =
-  \sum_{s=1}^{6}
I_5^{\mu_1 \dots \mu_{R-1} ,s } \bar{Q}_s^{\rho},
\eea
where the auxiliary vectors $\bar{Q}_s$ read:
\bea
 \bar{Q}_s^{\rho}&=&\sum_{i=1}^{6}  q_i^{\rho} \frac{{0s\choose 0i}_6}{{0\choose 0}_6}~~~,~~~ s=1 \dots 6.
\label{Q6}
\eea
Since ${00\choose 0i}_6=0$, the analogue of the first term in  (\ref{tensor5general}) with vector $\bar{Q}_0^{\rho}=\sum_{i=1}^{6}  q_i^{\mu} {00\choose 0i}_6 / {0\choose 0}_6$ vanishes. 
Equation (\ref{tensor6general}) is well-known 
\citep{Fleischer:1999hq,Binoth:2005ff,Denner:2005nn,Diakonidis:2008ij}. 
We only mention it in order to show how nicely it fits into our general scheme.
%--------------------------
\section{Discussion of the auxiliary vectors $Q_s^{\mu}$\label{sec-discussion}}
{Finally} we want to investigate properties of the {auxiliary} vectors {$Q_0^{\mu}, Q_s^{\mu}$ defined in}
 (\ref{Qs})  .
Of particular {practical} interest is the contraction {of a tensor integral} with  a chord.\footnote{{Remember that every external momentum may be expressed by the chords.}} 
{Usually,} a scalar product  {$q_i k$} is
expressed in terms of the difference of two propagators, which can then be cancelled 
such that a tensor integral of lower degree is obtained. 
This simplification {is often the} first step considering the original
diagram, where one may further select $q_n=0$.
Our approach offers an alternative
due to the fact that the contraction of one of the vectors
(\ref{Qs}) with 
a chord yields a simple expression (utilizing here the choice  $q_n=0$):
\bea
q_i  Q_0 &=& \sum_{j=1}^{n-1} q_i   q_j  \frac{{0\choose j}_n}{\left(  \right)_n} 
\nl
&=&
                -\frac{1}{2} \left(Y_{in}-Y_{nn} \right),~i=1, \cdots, n-1,
\label{sum1}
\\
q_i  Q_s &=& \sum_{j=1}^{n-1} q_i   q_j  \frac{{s\choose j}_n}{\left(  \right)_n} 
\nl
&=&
\frac{1}{2} \left({\delta}_{is}-{\delta}_{ns} \right),~i=1, \dots, n-1,
\nl && %\hspace*{2.5cm}
\phantom{\frac{1}{2} \left({\delta}_{is}-{\delta}_{ns} \right),}
{~s=1,\cdots, n,}
\label{sum2}
\eea 
with
\bea
\label{gram}
Y_{jk}=-(q_j-q_k)^2+m_j^2+m_k^2.
\eea
Further we observe that the contraction of any of the extra terms {introduced in (\ref{gmunun4}) and (\ref{gmunun3})} with a chord vanishes
 (for $q_n=0$).
In this way, we have  {not only} reduced
the rank of the tensor by one, {but we additionally}  obtain a simple expression in terms of the lower rank
tensor{s}. 

Relations (\ref{sum1}) and (\ref{sum2})  also allow a very simple derivation
of (\ref{tensor5general}) for $q_5=0$ by projecting with a complete set
of chords. 
Writing (\ref{tensor5general}) as:
\bea 
&&\int \frac{k^{\mu_1} \cdots k^{\mu_{R-1}} (k q_k)}{c_1 c_2 c_3 c_4 c_5} =
\int \frac{k^{\mu_1} \cdots k^{\mu_{R-1}} }{c_1 c_2 c_3 c_4 c_5} (Q_0 q_k) \nn \\
&&-\sum_{s=1}^{5} \int \frac{k^{\mu_1}  k^{\mu_{R-1}}\cdot c_s}{c_1 c_2 c_3 c_4 c_5}
(Q_s q_k),
\eea 
with $k q_k=-\frac{1}{2} \left[ c_k -c_5 +Y_{k5}-Y_{55} \right]$,
from (\ref{sum1}) and (\ref{sum2}) immediately follows  the equality.

For $q_5 \ne 0$ a shift, $k \rightarrow k+q_5$, is needed on the integration 
momentum $k$ of the original integral. Under such a shift all vectors $q_i$ are
shifted to $q_i \rightarrow q_i -q_5$. 
The  $Q_s$ for $s=1 \dots 5$ stay invariant, 
and the $Q_0$ shifts like a chord:
\bea
Q_0 = \sum_{j=1}^{5} q_j  \frac{{0\choose j}_5}{\left(  \right)_5} &\rightarrow&
\sum_{j=1}^{5} (q_j-q_5)  \frac{{0\choose j}_5}{\left(  \right)_5} 
\nonumber \\
&=&Q_0-q_5.
\eea 
This is due to the fact that the Gram determinants are invariant under any shift
and
\bea
\sum_{j=1}^{5}{0\choose j}_5&=&\left(  \right)_5,
\\
 \sum_{j=1}^{5}{s\choose j}_5&=&0 ~~,s=1 \dots 5.
\eea
After the shift one has only integrals with $q_5=0$. Collecting all contributions
and shifting back, one obtains (\ref{tensor5general}).

{The auxiliary vectors $Q_s$ introduce in the recursions inverse powers of $()_n$.}
{Relations} (\ref{sum1}) and (\ref{sum2}) yield:
\bea  
\label{eq09}
Q_0^2&=&\frac{1}{2}  \frac{{0\choose 0}_n}{\left(  \right)_n} + \frac{1}{2} Y_{nn} ,
\\
\label{eq10}
Q_s^2&=&\frac{1}{2}  \frac{{s\choose s}_n}{\left(  \right)_n}, {~s=1, \dots, n}.
\eea 
Equations
 {(\ref{eq09}) and (\ref{eq10}) {might suggest that after contractions with chords the effective inverse powers of $()_n$ are reduced.
This is not true in general.}
For $n=5$, however, this can be shown as follows:
We write  (\ref{sum1})
as a system of linear equations for the vector $Q_0^{\mu}$: 
\bea
\label{matrixequ}
A  Q_0 = x,
\eea
with
\bea
A&=&
\left( q_{i\mu} \right) ,
\\
x&=&
\begin{pmatrix}
   &x_1     & \\
   &\cdots    & \\
   &x_{4} &   
\end{pmatrix},
\\
x_i&=&-\frac{1}{2}(Y_{i5}-Y_{55}).
\eea
Obviously, the matrix $A$ satisfies
\bea
A_{i\mu} g^{\mu\nu} A_{j\nu} &=& q_i q_{j},~ i,j =1, \cdots 4,
\eea
and hence,
\bea
\mathrm{det}\left( q_i q_{j} \right)&=&-\frac{1}{2^{4}} \left( \right)_5
~=~ - \mathrm{det}^2\left(A\right) .
\eea
Therefore,
\bea
A^{-1} &\sim& { \left( \right)_5}^{-\frac{1}{2}}.
\eea
Solving  {the system} (\ref{matrixequ}) yields
\bea
\label{Q0beh}
Q_0&=&A^{-1} x ~\sim~ {\left( \right)_5}^{-\frac{1}{2}}.
\eea
{Thus the vector $Q_0$ is proportional to $()_5^{-1/2}$, while its additive
components in (\ref{Qs}) are proportional to $()_5^{-1}$.}
The same obviously applies for {the auxiliary vectors} $Q_s$ ($s=1 \dots 5$).
%%%%%%%%%%%%%%%%%%%%%%%%%%%%%%%%%%%%%%%%%%%%%%%%%%%%%%%%%%%%%%%%%%%%%%%%%%
\section*{Summary\label{sec-sum}}
%%%%%%%%%%%%%%%%%%%%%%%%%%%%%%%%%%%%%%%%%%%%%%%%%%%%%%%%%%%%%%%%%%%%%%%%%%
We have presented a new, recursive reduction scheme for one-loop $n$-point tensor Feynman integrals, and derived explicit expressions covering tensors up to rank $R=n$, with $n\leq 6$.
The crucial point is the derivation of relation (\ref{tensor5general}), which expresses $(5,R)$-integrals in terms of $(5,R-1)$- and $(4,R-1)$-integrals, and of relations like (\ref{tensor44}), which express  $(k,R)$-integrals ($k\leq 4$) in terms of $(k,R-1)$- and $(k-1,R-1)$-integrals plus additional terms.
Both types of representations make use of auxiliary vectors $Q_s^{\mu}$.

The   {recursive} scheme is very convenient for explicit calculations and contains the complete calculational chain of tensor reduction, for both massive and massless propagators, and works with dimensional regularization.
The only necessary package to be added is one for the evaluation of
1-point to 4-point  scalar integrals.
For this we chose the library of scalar functions QCDloop/FF \citep{Ellis:2007qk,vanOldenborgh:1990yc}
and made careful numerical checks of the recursive reductions.
The numerical output was compared,  whenever possible, to another, independent  Fortran program \citep{Diakonidis:2008ij,Diakonidis:2008dt}.
Additionally, 6-point functions have been compared with a
program of P. Uwer for massive particles \citep{Uwer-privcommun:2009}, and 5- and 6-point functions with the Golem package \citep{Binoth:2008uq}, which applies for massless particles only. 
The highest rank covered was $R=5$.
We have agreement in all cases, both for the divergent and finite parts of the tensor integrals.
  {Various  tables with} numerical comparisons may be found in \citep{Diakonidis-calc:2009}.

Differing from the reductions derived in \citep{Diakonidis:2008ij,Diakonidis:2008dt},   {one cannot} avoid the appearance of inverse Gram determinants $()_5$   {in the recursions}.
There is a lengthy discussion in the literature, and the rising numerical experience with explicit calculations of cross-sections leads to the conclusion that it is a small fraction of phase-space points which is substantially concerned. 
For those points,   {we have to switch to our code based on \citep{Diakonidis:2008ij,Diakonidis:2008dt}, which is more appropriate to the specific kinematics around $()_5=0$.}
   {
But doing so, the recursive character of the scheme is  given up.} 
It is also possible, in the recursion described here, to re-introduce and/or keep integrals in higher dimension in order to avoid inverse Gram determinants, as exemplified in another approach \citep{Binoth:2005ff}.

%  

%%%%%%%%%%%%%%%%%%%%%%%%%%%%%%%%%%%%%%%%%%%%%%%%%%%%%%%%%%%%%%%%%%%%%%%%%%
%\clearpage
\section*{Acknowledgments}
%%%%%%%%%%%%%%%%%%%%%%%%%%%%%%%%%%%%%%%%%%%%%%%%%%%%%%%%%%%%%%%%%%%%%%%%%%
Work supported in part by Sonderforschungsbereich/Trans\-re\-gio
 SFB/\linebreak[3]TRR 9 of DFG 
``Com\-pu\-ter\-ge\-st\"utz\-te Theoretische Teil\-chen\-phy\-sik"
and by the European Community's Marie-Curie Research Trai\-ning Network
MRTN-CT-2006-035505 
``HEP-\linebreak[4]TOOLS''. %%%%%% the name is: HEPTOOLS
J.F. likes to thank DESY for kind hospitality. 

% \bibliographystyle{elsarticle-num} % bst file from elsevier 2009
% \bibliography{2loops} %%%%%%%%%%%%,2loops_teo}

\begin{thebibliography}{10}
\expandafter\ifx\csname url\endcsname\relax
  \def\url#1{\texttt{#1}}\fi
\expandafter\ifx\csname urlprefix\endcsname\relax\def\urlprefix{URL }\fi
\expandafter\ifx\csname href\endcsname\relax
  \def\href#1#2{#2} \def\path#1{#1}\fi

\bibitem{Passarino:1978jh}
G.~Passarino, M.~J.~G. Veltman, One loop corrections for $e^+ e^-$ annihilation
  into $\mu^{+} \mu^{-}$ in the {Weinberg} model, Nucl. Phys. B160 (1979) 151.
\newblock \href {http://dx.doi.org/10.1016/0550-3213(79)90234-7}
  {\path{doi:10.1016/0550-3213(79)90234-7}}.

\bibitem{Weinzierl:2007vk}
S.~Weinzierl, {Automated calculations for multi-leg processes}, PoS ACAT (2007)
  005.
\newblock \href {http://arxiv.org/abs/0707.3342} {\path{arXiv:0707.3342}}.


%\cite{Bern:2008ef}
\bibitem{Bern:2008ef}
  Z.~Bern et al.  [NLO Multileg Working Group],
  The NLO multileg working group: Summary report.
%  arXiv:0803.0494 [hep-ph].
  %%CITATION = ARXIV:0803.0494;%%
\newblock \href {http://arxiv.org/abs/0803.0494} {\path{arXiv:hep-ph/0803.0494}}.

\bibitem{Binoth:2009fk}
T.~Binoth, {LHC phenomenology at next-to-leading order QCD: theoretical
  progress and new results}, PoS(ACAT08) 011.
\newblock \href {http://arxiv.org/abs/0903.1876} {\path{arXiv:0903.1876}}.

\bibitem{Diakonidis:2008ij}
T.~Diakonidis, J.~Fleischer, J.~Gluza, K.~Kajda, T.~Riemann, J.B.~Tausk,  {A complete reduction of one-loop tensor 5- and 6-point integrals}, Phys. Rev. D80 (2009) 036003.
\newblock \href {http://arxiv.org/abs/0812.2134} {\path{arXiv:0812.2134}}.
%  \href {http://dx.doi.org/10.1103/PhysRevD.80.036003} {\path{doi:10.1103/PhysRevD.80.036003}}.

\bibitem{Davydychev:1991va}
A.~I. Davydychev, {A Simple formula for reducing Feynman diagrams to scalar
  integrals}, Phys. Lett. B263 (1991) 107--111.
\\
\newblock \href {http://dx.doi.org/10.1016/0370-2693(91)91715-8}
  {\path{doi:10.1016/0370-2693(91)91715-8}}.

\bibitem{Tarasov:1996br}
O.~V. Tarasov, Connection between {Feynman} integrals having different values
  of the space-time dimension, Phys. Rev. D54 (1996) 6479--6490.
\newblock \href {http://arxiv.org/abs/hep-th/9606018}
  {\path{arXiv:hep-th/9606018}}.

\bibitem{Melrose:1965kb}
D.~B. Melrose, Reduction of {Feynman} diagrams, Nuovo Cim. 40 (1965) 181--213.

\bibitem{Fleischer:1999hq}
J.~Fleischer, F.~Jegerlehner, O.~Tarasov, Algebraic reduction of one-loop
  {Feynman} graph amplitudes, Nucl. Phys. B566 (2000) 423--440.
\\
\newblock \href {http://arxiv.org/abs/hep-ph/9907327}
  {\path{arXiv:hep-ph/9907327}}.

\bibitem{Diakonidis:2008dt}
T.~Diakonidis, J.~Fleischer, J.~Gluza, K.~Kajda, T.~Riemann, J.B.~Tausk, {On the
  tensor reduction of one-loop pentagons and hexagons}, Nucl. Phys. Proc.
  Suppl. 183 (2008) 109--115.
\newblock \href {http://arxiv.org/abs/0807.2984} {\path{arXiv:0807.2984}}.
%\\
%  \href {http://dx.doi.org/10.1016/j.nuclphysbps.2008.09.091} {\path{doi:10.1016/j.nuclphysbps.2008.09.091}}.

\bibitem{Hahn:1998yk2}
T.~Hahn, M.~Perez-Victoria, Automatized one-loop calculations in four and d
  dimensions, Comput. Phys. Commun. 118 (1999) 153--165.
\\
\newblock \href {http://arxiv.org/abs/hep-ph/9807565}
  {\path{arXiv:hep-ph/9807565}}.

\bibitem{vanOldenborgh:1990yc}
G.~J. van Oldenborgh, {FF: A Package to evaluate one loop Feynman diagrams},
  Comput. Phys. Commun. 66 (1991) 1--15.
\\
\newblock \href {http://dx.doi.org/10.1016/0010-4655(91)90002-3}
  {\path{doi:10.1016/0010-4655(91)90002-3}}.

\bibitem{Binoth:2008uq}
T.~Binoth, J.~P. Guillet, G.~Heinrich, E.~Pilon, T.~Reiter, {Golem95: a
  numerical program to calculate one-loop tensor integrals with up to six
  external legs},
Comput. Phys. Commun.  {180} (2009) 2317.
\\
 \href {http://arxiv.org/abs/0810.0992}
  {\path{arXiv:0810.0992}}.


\bibitem{Binoth:2005ff}
T.~Binoth, J.~Guillet, G.~Heinrich, E.~Pilon, C.~Schubert, An algebraic /
  numerical formalism for one-loop multi-leg amplitudes, JHEP 10 (2005) 015.
\newblock \href {http://arxiv.org/abs/hep-ph/0504267}
  {\path{arXiv:hep-ph/0504267}}.

%\cite{Denner:2002ii}
\bibitem{Denner:2002ii}
  A.~Denner and S.~Dittmaier,
  Reduction of one-loop tensor 5-point integrals,
  Nucl. Phys.  B658 (2003) 175.
\newblock \href {http://arxiv.org/abs/hep-ph/0212259}
  {\path{arXiv:hep-ph/0212259}}.
  %%CITATION = NUPHA,B658,175;%%


\bibitem{Denner:2005nn}
A.~Denner, S.~Dittmaier, Reduction schemes for one-loop tensor integrals, Nucl.
  Phys. B734 (2006) 62--115.
\newblock \href {http://arxiv.org/abs/hep-ph/0509141}
  {\path{arXiv:hep-ph/0509141}}.

%\cite{delAguila:2004nf}
\bibitem{delAguila:2004nf}
  F.~del Aguila and R.~Pittau,
  Recursive numerical calculus of one-loop tensor integrals,
  JHEP 0407 (2004) 017.
\newblock \href {http://arxiv.org/abs/hep-ph/0404120}
  {\path{arXiv:hep-ph/0404120}}.
%  [arXiv:hep-ph/0404120].
  %%CITATION = JHEPA,0407,017;%%


%\cite{vanHameren:2009vq}
\bibitem{vanHameren:2009vq}
  A.~van Hameren,
  Multi-gluon one-loop amplitudes using tensor integrals,
  JHEP 0907 (2009) 088.
%  [arXiv:0905.1005 [hep-ph]].
  %%CITATION = JHEPA,0907,088;%%
\newblock \href {http://arxiv.org/abs/hep-ph/0905.1005}
  {\path{arXiv:hep-ph/0905.1005}}.

\bibitem{vanNeerven:1983vr}
W.~L. van Neerven, J.~A.~M. Vermaseren, {Large loop integrals}, Phys. Lett.
  B137 (1984) 241.
%\\
\newblock \href {http://dx.doi.org/10.1016/0370-2693(84)90237-5}
  {\path{doi:10.1016/0370-2693(84)90237-5}}.

\bibitem{Bern:1992em}
Z.~Bern, L.~J. Dixon, D.~A. Kosower, {Dimensionally Regulated One-Loop
  Integrals}, Phys. Lett. B302 (1993) 299--308 [Erratum--ibid.\ B {\bf 318}
  (1993) 649].
\newblock \href {http://arxiv.org/abs/hep-ph/9212308}
  {\path{arXiv:hep-ph/9212308}}.
% \href
%   {http://dx.doi.org/10.1016/0370-2693(93)90400-C}
%   {\path{doi:10.1016/0370-2693(93)90400-C}}.

\bibitem{Bern:1993kr}
Z.~Bern, L.~J. Dixon, D.~A. Kosower, {Dimensionally regulated pentagon
  integrals}, Nucl. Phys. B412 (1994) 751--816.
\\
\newblock \href {http://arxiv.org/abs/hep-ph/9306240}
  {\path{arXiv:hep-ph/9306240}}.
%  \href
%   {http://dx.doi.org/10.1016/0550-3213(94)90398-0}
%   {\path{doi:10.1016/0550-3213(94)90398-0}}.

\bibitem{Binoth:1999sp}
T.~Binoth, J.~P. Guillet, G.~Heinrich, {Reduction formalism for dimensionally
  regulated one-loop N-point integrals}, Nucl. Phys. B572 (2000) 361--386.
\newblock \href {http://arxiv.org/abs/hep-ph/9911342}
  {\path{arXiv:hep-ph/9911342}}.
%  \href
%   {http://dx.doi.org/10.1016/S0550-3213(00)00040-7}
%   {\path{doi:10.1016/S0550-3213(00)00040-7}}.

\bibitem{Duplancic:2003tv}
G.~Duplancic, B.~Nizic, {Reduction method for dimensionally regulated one-loop
  N- point Feynman integrals}, Eur. Phys. J. C35 (2004) 105--118.
\newblock \href {http://arxiv.org/abs/hep-ph/0303184}
  {\path{arXiv:hep-ph/0303184}}.
%  \href
%   {http://dx.doi.org/10.1140/epjc/s2004-01723-7}
%   {\path{doi:10.1140/epjc/s2004-01723-7}}.



\bibitem{vanOldenborgh:1989wn}
G.~J. van Oldenborgh, J.~A.~M. Vermaseren, {New Algorithms for One Loop
  Integrals}, Z. Phys. C46 (1990) 425--438.
\\
\newblock \href {http://dx.doi.org/10.1007/BF01621031}
  {\path{doi:10.1007/BF01621031}}.

% \bibitem{Fleischer-calc:2009}
% J. Fleischer, Talk at CALC 2009, July 2009, Dubna, Russia.
% \newblock \href{http://theor.jinr.ru/~calc2009/program.html}{[link]}.
% \newline\urlprefix\url{http://theor.jinr.ru/~calc2009/program.html}

\bibitem{Fleischer-calc:2009}
J. Fleischer, Talk at CALC 2009, July 2009, Dubna, Russia.
Transparencies at
%\\
\newblock \href{http://theor.jinr.ru/~calc2009/program.html}{http://theor.jinr.ru/~calc2009/program.html}.
%  http://theor.jinr.ru/~calc2009/talk/fleischer.pdf.

\bibitem{Ellis:2007qk}
R.~K. Ellis, G.~Zanderighi, {Scalar one-loop integrals for QCD}, JHEP 02 (2008)
  002.
\newblock \href {http://arxiv.org/abs/0712.1851} {\path{arXiv:0712.1851}}.
% 
%   \href {http://dx.doi.org/10.1088/1126-6708/2008/02/002}
%   {\path{doi:10.1088/1126-6708/2008/02/002}}.

\bibitem{Uwer-privcommun:2009}
P.~Uwer, private communication.

% \bibitem{Diakonidis-calc:2009}
% T. Diakonidis, Talk at CALC 2009, July 2009, Dubna, Russia.
% \newblock \href{http://theor.jinr.ru/~calc2009/program.html}{[link]}.
% \newline\urlprefix\url{http://theor.jinr.ru/~calc2009/program.html}

\bibitem{Diakonidis-calc:2009}
T. Diakonidis, Talk at CALC 2009, July 2009, Dubna, Russia. Transparencies at
%\\
\newblock \href{http://theor.jinr.ru/~calc2009/program.html}{http://theor.jinr.ru/~calc2009/program.html}.
%\newline\urlprefix\url{http://theor.jinr.ru/~calc2009/program.html}


\end{thebibliography}
% 
% 
% \end{document}

\end{document}